\documentclass[lettersize,journal]{IEEEtran}
\usepackage{amsmath,amsfonts, textgreek}
\usepackage{algorithmic}
\usepackage{algorithm}
\usepackage{array}
\usepackage[caption=false,font=normalsize,labelfont=sf,textfont=sf]{subfig}
\usepackage{textcomp}
\usepackage{stfloats}
\usepackage{url}
\usepackage{verbatim}
\usepackage{graphicx}
\usepackage{cite}
\usepackage{orcidlink}
\usepackage{aas_macros}

\begin{document}

\title{Optimization of lenslet arrays for PRIMA Kinetic Inductance Detectors}

\author{Sumit Dahal\orcidlink{0000-0002-1708-5464},
        Thomas R. Stevenson\orcidlink{0000-0001-6221-2357}, Nicholas P. Costen, Nat DeNigris, Jason Glenn\orcidlink{0000-0001-7527-2017}, Gang Hu, Christine A. Jhabvala, Ricardo Morales-Sanchez, Jessica B. Patel, Manuel A. Quijada, Ian Schrock, Frederick H. Wang, and Edward J. Wollack\orcidlink{0000-0002-7567-4451}
\thanks{Manuscript submitted: 13 August 2025; Date of publication: 6 April 2026  (\textit{Corresponding author: S. Dahal})}
\thanks{S. Dahal is with NASA Goddard Space Flight Center, Greenbelt, MD, USA and also with Johns Hopkins University, Baltimore, MD, USA (e-mail: sumit.dahal@nasa.gov).}
\thanks{T. R. Stevenson, N. P. Costen, N. DeNigris, J. Glenn, G. Hu, C. A. Jhabvala, R. Morales-Sanchez, J. B. Patel, M. A. Quijada, I. Schrock, F. H. Wang, and E. J. Wollack are with NASA Goddard Space Flight Center, Greenbelt, MD, USA.}
}

\markboth{IEEE TRANSACTIONS ON APPLIED SUPERCONDUCTIVITY,~Vol.~36, No.~6, September~2026}%
{Shell \MakeLowercase{\textit{et al.}}: A Sample Article Using IEEEtran.cls for IEEE Journals}


\maketitle

\begin{abstract}
The PRobe far-Infrared Mission for Astrophysics (PRIMA) is a cryogenically cooled 1.8-m space telescope designed to address fundamental questions about the evolution of galactic ecosystems, the origins of planetary atmospheres, and the buildup of dust and metals over cosmic time. PRIMA will achieve unprecedented sensitivity in the 24 -- 261 \textmu  m wavelength range, enabled by background-limited kinetic inductance detectors (KIDs) cooled to 120 mK. For PRIMA’s Far-InfraRed Enhanced Survey Spectrometer (FIRESS) instrument, we have developed monolithic kilopixel silicon lenslet arrays to efficiently couple incident radiation from the telescope’s fore-optics onto the KID absorber elements. These three-dimensional lenslet arrays are fabricated using grayscale lithography, followed by deep reactive ion etching (DRIE), and are anti-reflection (AR) coated with a quarter-wavelength thick deposition of \mbox{Parylene-C}. The lenslet arrays are aligned and bonded to the KID arrays using a thin layer of epoxy through a flip-chip bonder. In this work, we report on the optimized fabrication, lens design, AR coating, and bonding processes developed for the FIRESS lenslet arrays. We characterize brassboard lenslet arrays fabricated to meet the specifications of the FIRESS low and high spectral bands, demonstrate stepped-thickness AR-coatings to achieve high efficiency across broad wavelength ranges, and present spectral transmission measurements of the AR coating and the epoxy bonding layers.
\end{abstract}


\section{Introduction}\label{sec:intro}

\begin{figure}[t]
\centering
\includegraphics[width=3.4in, trim={2.2in 0.5in 1.6in 2.3in},clip]{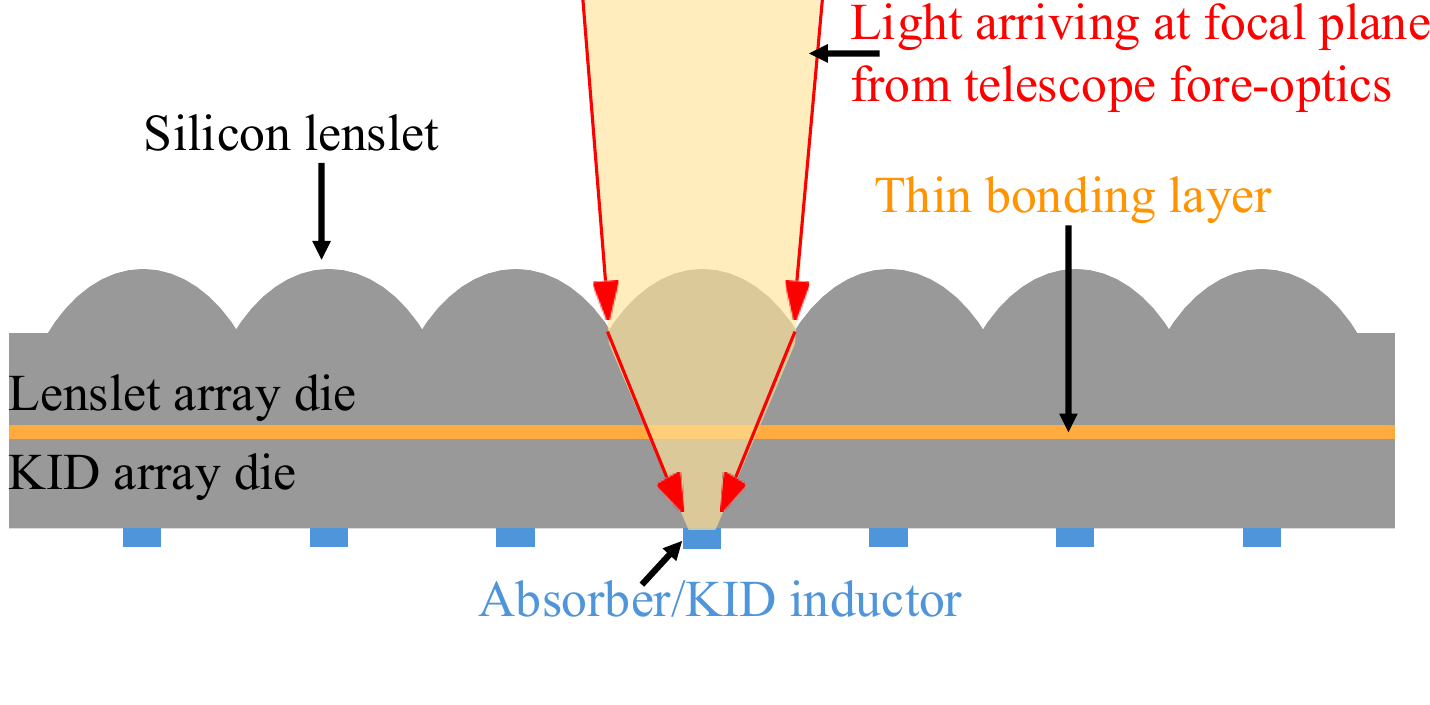}
\includegraphics[width=3.4in]{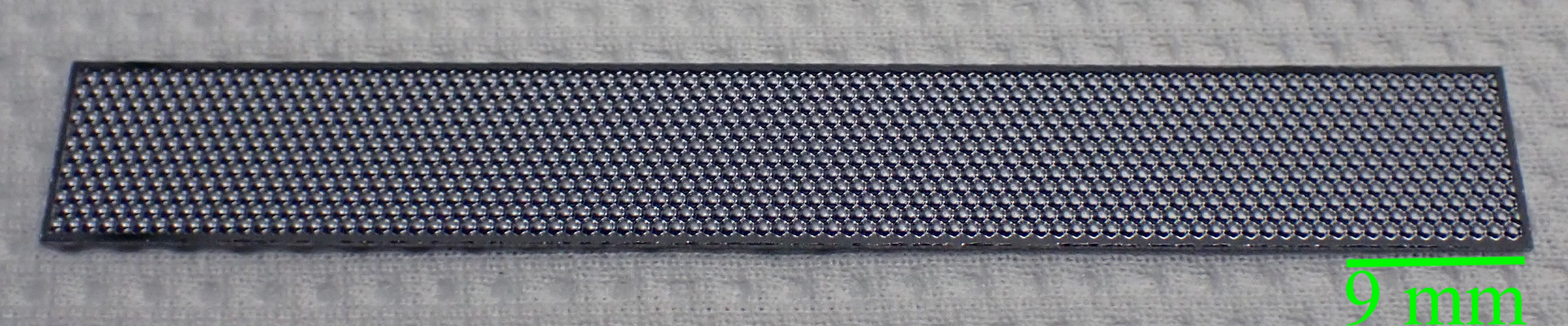}
\caption{Top: Illustration of PRIMA lenslet-detector array configuration. The lenslets focus the incoming radiation from the telescope fore-optics onto the KID absorber elements. The lenslet and the detector arrays are fabricated on separate silicon substrates and then bonded together with a thin adhesive layer. Updated figure from \cite{cothard2024}. Bottom: Photograph of a PRIMA FIRESS long waveband lenslet array. The 10.613 mm wide $\times$ 77.05 mm long die consists of 12 $\times$ 84 lenslets, hexagonally packed with a 900 \textmu m pitch.}
\label{fig:lens_config}
\end{figure}

\IEEEPARstart{T}{he} PRobe far-Infrared Mission for Astrophysics (PRIMA) \cite{glenn2025} is an observatory concept proposed as NASA's probe-class mission to address fundamental questions about the evolution of galactic ecosystems, the origins of planetary atmospheres, and the buildup of dust and metals over cosmic time. With most of its observing time (75\%) dedicated to general observer programs, PRIMA opens new discovery space in the 24 -- 261 \textmu m wavelength range, bridging the gap between existing infrared and radio observatories. PRIMA features a cryogenically cooled 1.8-m telescope with $\sim$ 11,000 background-limited kinetic inductance detectors (KIDs) \cite{day2024, yates2025}, providing unprecedented sensitivity at these wavelengths. To achieve such high sensitivities, the KIDs on the detector focal plane must be efficiently coupled to the incident radiation from the telescope's fore-optics. In \cite{cothard2024}, we discussed the development of monolithic silicon lenslet arrays suitable for optical coupling to far-infrared detectors. This paper presents follow-up work optimizing kilo-pixel lenslet arrays for PRIMA KIDs. Figure \ref{fig:lens_config} shows an image of one of the fabricated PRIMA kilo-pixel arrays and an illustration of the lenslets coupling incoming radiation to KID absorber elements (which also act as inductors for the microwave resonators).

PRIMA consists of two science instruments: the Far-InfraRed Enhanced Survey Spectrometer (FIRESS) \cite{bradford2025} -- a wide-band spectrometer, and the PRIMA Imager (PRIMAger) \cite{ciesla2025} -- a multi-band spectrophotometric imager/polarimeter. In this work, we focus on lenslet optimization for the FIRESS instrument. Although PRIMAger prototype detector arrays use commercially obtained laser-machined silicon lens arrays from Veldlaser \cite{baselmans2022}, we note that the techniques and strategies described herein can be applied to develop lenslets for PRIMAger or other far-infrared instruments. We describe the fabrication process for the FIRESS lenslet arrays in Section \ref{sec:fab}. In Section \ref{sec:lens_design}, we present the lens design and profile measurements, demonstrating that the improved fabrication process meets the FIRESS requirements. Sections~\ref{sec:ar_coating} and \ref{sec:bonding} report on the stepped-thickness anti-reflection (AR) coating and the bonding to the detector array, respectively. Finally, we provide a summary and discuss the next steps in Section \ref{sec:summary}.

\section{Fabrication}\label{sec:fab}
At millimeter and sub-millimeter wavelengths, several detector optical coupling schemes including feedhorn arrays \cite{simon2018, liu2022}, individual hemispherical lenslets \cite{Siritanasak2016, benson2014, Taniguchi2022}, and gradient index lenses \cite{defrance2020, pisano2020} have been successfully demonstrated. However, extending these technologies to shorter far-infrared wavelengths is challenging due to more stringent requirements on lens profile accuracy, surface roughness, and detector-lens alignment. In our earlier work \cite{cothard2024}, we demonstrated that grayscale lithography followed by deep reactive ion etching (DRIE) is a viable method to fabricate lenslet arrays for the entire \mbox{24 -- 261 \textmu m} PRIMA wavelength range. We use grayscale lithography to pattern a thick photoresist layer with a surface designed to produce a lens array with desired profiles in silicon after an etch process. 

In this work, we focus on optimizing the shortest (24 -- 43 \textmu m) and the longest (130 -- 235 \textmu m) FIRESS wavebands, referred to hereafter as ``Band~1'' and ``Band~4,'' respectively. The optimizations for the two intermediate FIRESS wavebands are expected to be relatively straightforward interpolations from our Band~1 and Band~4 designs. Being at the extreme ends of the FIRESS wavelength range, Band~1 and Band~4 present unique challenges for lenslet fabrication and bonding. For Band~1, the primary challenge was reducing the epoxy bond layer thickness (see Section \ref{sec:bonding}) to lower loss of optical efficiency from reflection and absorption. We achieved this by redesigning our Band~1 lenses to use a thicker silicon die, which is discussed in detail in Section \ref{sec:lens_design}. 

For Band~4, the challenge was in etching lenses deep enough in a thin die to be optically fast enough to focus light on the small detector absorber. In our prior work \cite{cothard2024}, we etched Band~4 arrays with a lens profile close to design up to a depth of 145 \textmu m at a round perimeter of 826 \textmu m. The deeper etching was limited by the distortion of the lens profile due to a lateral etching effect. Our etch process for grayscale transfer (patent pending) differs from prior art \cite{morgan2004, morgan2005a, morgan2005b} in that we omit a DRIE passivation polymer formation step. Instead, we use alternating cycles of (1) silicon etch in fluorine plasma, and (2) resist etch in oxygen plasma. With this approach, we are able to etch smooth, deeply curved silicon surfaces with any selectivity we desire within a broad range. However, as we described in \cite{cothard2024}, the deviation at large radii and depths we encountered in fabricating Band~4 lenses was due to the etch being partly isotropic, causing a significant lateral etch at the steepest portions of the lens.  For the new Band 4 lenses described here (see Figure \ref{fig:hex_lens_design}), we obtained a good fit to our etch data using a model in which 65 -- 75\% of the total etch rate is from an isotropic etch process, with the remainder from a fully anisotropic (straight down) etch process. We then designed an initial resist profile to compensate for this effect. 

To achieve a desired final silicon lens depth at a particular radius, one needs to start with a resist profile that has its depth modified at a larger radius, allocating room for lateral etch correction. At the locations where the lenses ideally touch the edges of the hexagonal unit cells, i.e. at inradius ($R_i$) of 450~\textmu m, there is no such room available in the array. However, in the direction of the hexagonal cell corners at circumradius ($R_c = 2/\sqrt{3} \ R_i$) of $\approx$~520 \textmu m, the lenses can be made deeper with a lateral etch correction out to larger radii than the inradius of the hexagonal cells. As a result, we can improve the light collecting area of the lenses in the hexagonal corners. The Band 4 lenslet arrays with hexagonal corners made through this enhanced fabrication process is presented in Section \ref{sec:lens_design}.

\section{Lens Design}\label{sec:lens_design}
Each FIRESS lenslet array consists of 1008 pixels (12 spatial $\times$ 84 spectral), hexagonally packed with a pitch of 900~\textmu m, as shown in Figure \ref{fig:lens_config}. The lenslets are designed to focus light onto the KID absorbers that are much smaller in size, ranging from \mbox{70 -- 115 \textmu m} in diameter \cite{day2024}. The lens design is optimized individually for each of the four FIRESS wavebands. Here, we present design optimizations for FIRESS Band~1 and Band~4 arrays. 

\begin{figure}[t]
\centering
\includegraphics[width=3.4in, trim={0.75in 0.5in 0.85in 1in},clip]{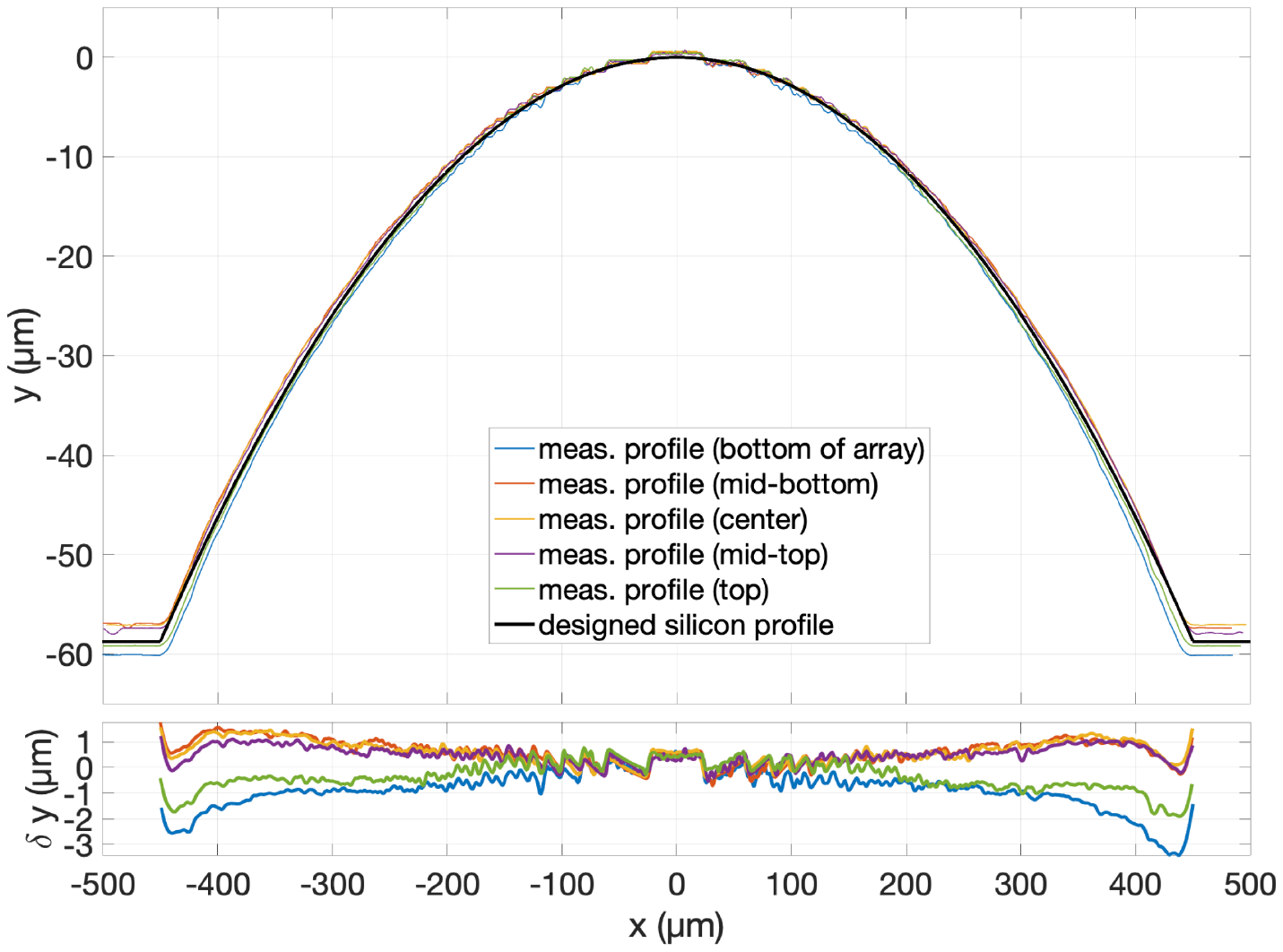}
\caption{Profilometer measurements of a FIRESS Band 1 array. The top plot compares the design with the measured lens profiles from five different regions of the array. The corresponding residuals in the bottom plot show RMS error \mbox{$<$ 0.9 \textmu m} for all regions, except 1.5 \textmu m for the bottom of the array. The fabrication tolerances achieved within \mbox{$\sim$ 375 \textmu m} radius in our prior work (see Figure~2 in \cite{cothard2024}) have now been extended up to \mbox{$\sim$ 420 \textmu m}, highlighting the improvement in our fabrication process.}
\label{fig:lens_design}
\end{figure}

In our earlier work presented in \cite{cothard2024}, we fabricated Band~4 lenslet arrays on a custom 525 \textmu m thick high-resistivity silicon wafer. For Band~1, the dies must withstand higher bond forces to minimize the thickness of the epoxy bond layer between the lenslet and the detector (see Section \ref{sec:bonding}). Therefore, we use a thicker (1560 \textmu m) die for Band~1 to maximize mechanical stiffness. The use of a thicker die also reduces the required lens curvature, enabling better lens profile accuracy during fabrication, as highlighted in Figure \ref{fig:lens_design}. For the 900~\textmu m lens diameter, our chosen die thicknesses allow diffraction-limited spot sizes of $\approx$ 26 and 61 \textmu m diameters for Bands~1 and ~4, respectively. Both of these spot sizes are smaller than their respective detector absorber diameters of 70 and 115 \textmu m.

\begin{figure}[t]
\centering
\includegraphics[width=3.4in]{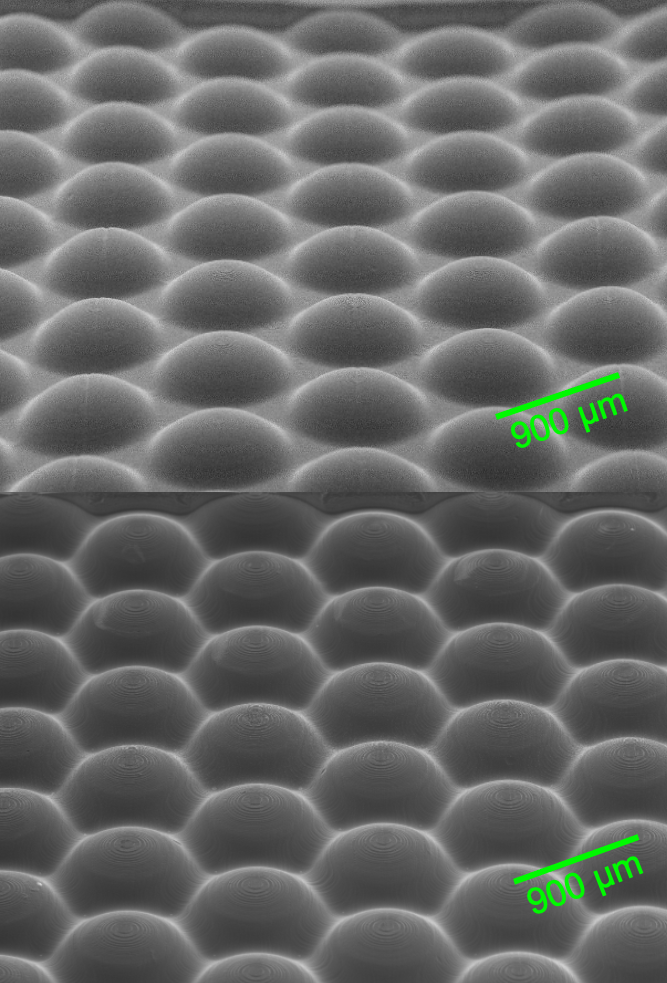}
\includegraphics[width=3.4in]{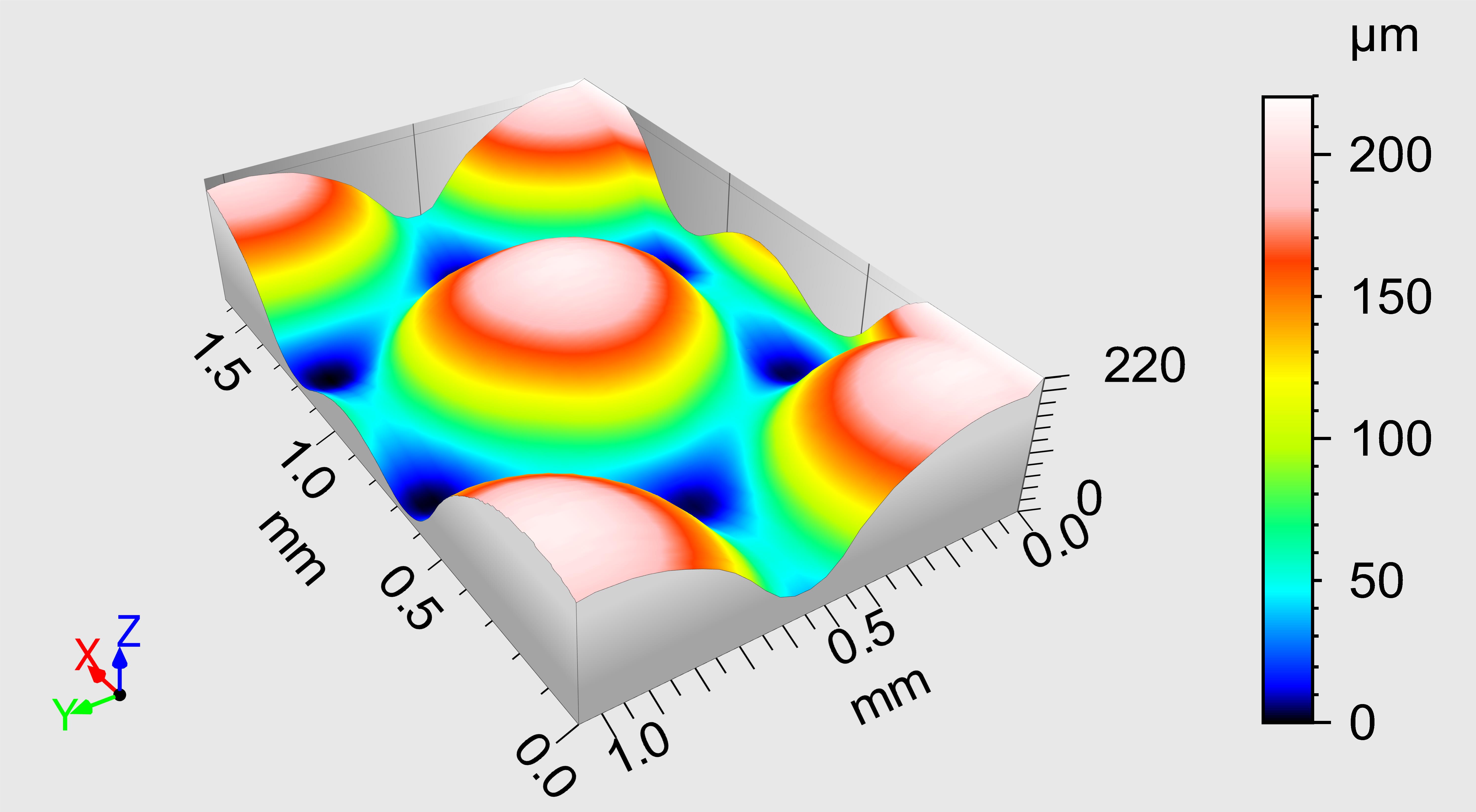}
\caption{Top: Scanning electron microscope (SEM) image of a FIRESS Band~4 array from our prior work \cite{cothard2024}. Middle: SEM image of an upgraded Band~4 array with reduced flat spaces between the lenses for increasing light collecting area and controlling stray light. The concentric steps visible towards the center of the lenses are due to discrete grayscale exposure levels. Bottom: 3D profilometry of one of the lenses from the upgraded array. With a resist profile compensated for lateral etching effect (see Section \ref{sec:fab}), the lenses could be etched deep enough to produce the desired profile with hexagonal pockets.}
\label{fig:hex_lens_design}
\end{figure}

It is not desirable to focus perfectly on the absorber, as this could result in spot sizes similar to or smaller than the optically resonant structure of the absorber ($\approx$ 4 \textmu m). Consequently, our prior lenses \cite{cothard2024} were designed to focus slightly past the detector plane to provide more uniform illumination of the absorber. However, variations in the curvatures of the fabricated lenses (see Figure \ref{fig:lens_design}) could lead to larger defocusing, reducing optical efficiency. Therefore, our current baseline design does not include defocusing so that our tolerance on the lens figure remains within the depth of focus. The effect on the optical beam from such offsets within the depth of focus is negligible. There are larger effects on the beam size and shape that result from our sampling at the focal plane, which is described in \cite{bradford2025} and analyzed in \cite{Berkel2025}. 

The lenses are hexagonally packed in the kilo-pixel array to match the 900 \textmu m pixel spacing of the KID absorbers on the detector array. To develop our fabrication process and demonstrate its viability for FIRESS Bands 1 and 4, we used circular lenses as shown in Figure \ref{fig:lens_design}. These lenses do not extend into the corners of hexagonal unit cell, leading to flat areas between them as shown in the top image of Figure \ref{fig:hex_lens_design}. These ``dead spaces'' do not focus light onto KID absorbers and could contribute to scattering within the bonded lenslet-detector array, leading to stray light. To increase the light collecting area, and hence the optical efficiency, we re-designed the lenses to increase the overlap in the hexagonal corners as shown in the middle and the bottom plots of Figure~\ref{fig:hex_lens_design}. This was made possible by our enhanced fabrication process corrected for lateral etching (Section \ref{sec:fab}), allowing deeper etch at larger radii. Following the lens encircled power model presented in \cite{cothard2024}, we estimate that the upgraded Band~4 lenses with hexagonal corners (Figure \ref{fig:hex_lens_design}, middle/bottom) direct $\approx$~14\% more optical power to the detectors than the prior lenses with circular profile (Figure \ref{fig:hex_lens_design}, top). We plan to design and fabricate lenses with hexagonal corners for the remaining three FIRESS bands as well.  

\section{Anti-Reflection Coating}\label{sec:ar_coating}
\begin{figure}[ht]
\centering
\includegraphics[width=3.4in]{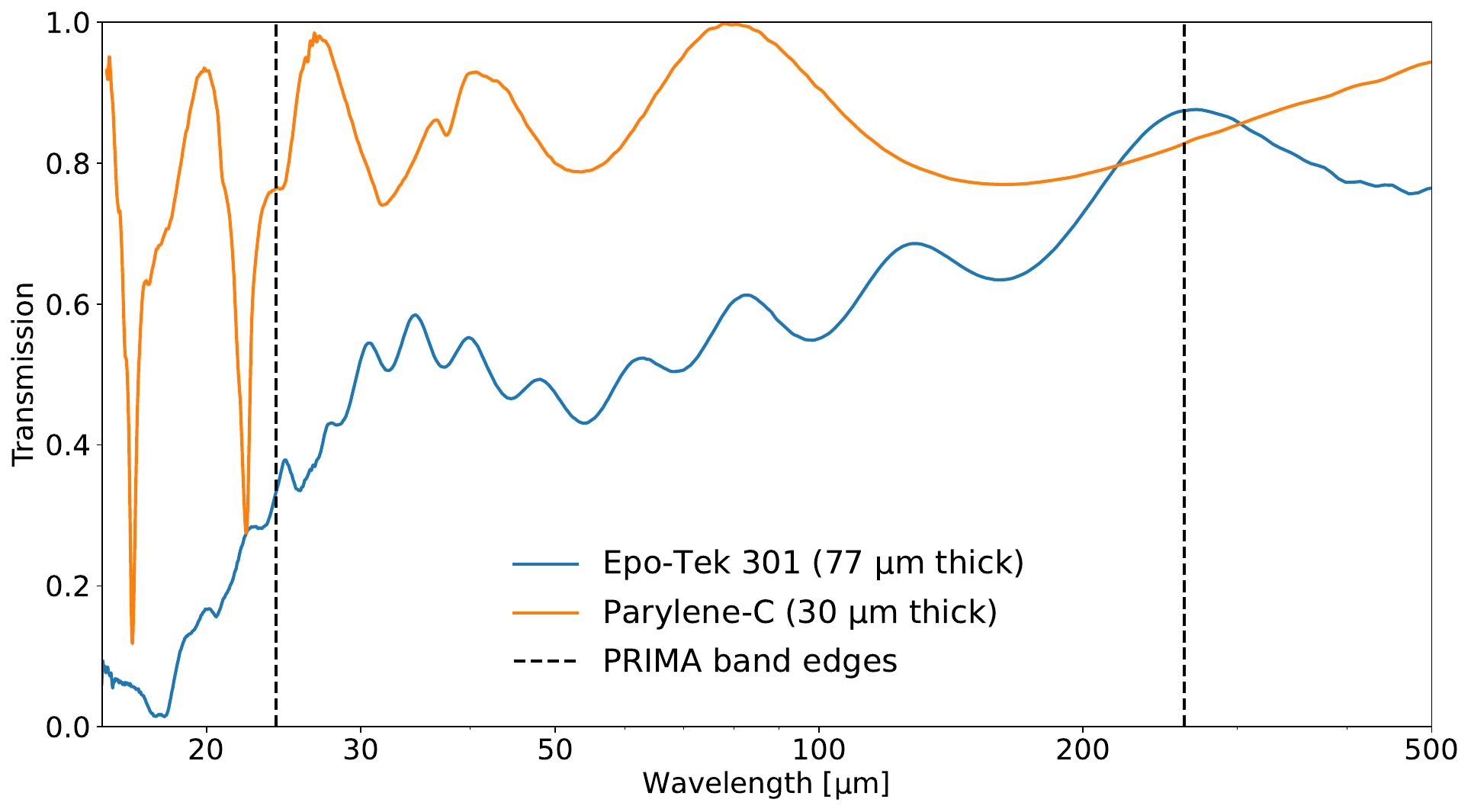}
\caption{Measured spectral transmission for free-standing Epo-Tek 301 and Parylene-C samples cooled to 5 K. These data were used to extract the complex dielectric function of the materials at PRIMA wavelengths to optimize the bonding and the AR-coating thicknesses. The actual thicknesses (see Sections \ref{sec:ar_coating} and \ref{sec:bonding}) used for PRIMA lenslets are smaller, providing correspondingly higher transmission than the thicker free-standing samples shown here for optical characterization of the materials.}
\label{fig:fts}
\end{figure}

The lenslet arrays are AR-coated with a quarter-wavelength thick Parylene-C using an SCS Labcoter Parylene deposition system\footnote{www.scscoatings.com/parylene-coatings/}. For a given wavelength ($\lambda$), the desired coating thickness is given by $d = \lambda/(4n)$, where $n$ is the refractive index of the medium. To determine $n$ at the far-infrared PRIMA wavelengths, we prepared a 30 \textmu m thick free-standing Parylene-C sample and characterized it using a Fourier transform spectrometer (FTS). The spectral transmission through the sample at 5 K (liquid helium bath temperature) is shown in Figure \ref{fig:fts}. The two prominent absorption lines lie safely below the shortest PRIMA band edge. These spectrally resolved transmission data were fit to a model to extract the material's complex dielectric function, which will be described in detail in an upcoming publication (Wollack et. al., in prep.). The FTS measurements with the sample at 5 K bath temperature represent the material's expected optical properties (i.e. the complex dielectric function) at the PRIMA focal plane operating temperature of 120 mK as well. At PRIMA wavelengths, we determined $1.66 < n(\lambda) < 1.68$. This corresponds to AR-coating thicknesses of 5 \textmu m and 28 \textmu m for the FIRESS Band~1 and Band~4 centers, respectively.

\begin{figure}[t]
\centering
\includegraphics[width=2.38in, trim={1.5in 0in 1.5in 1.5in},clip]{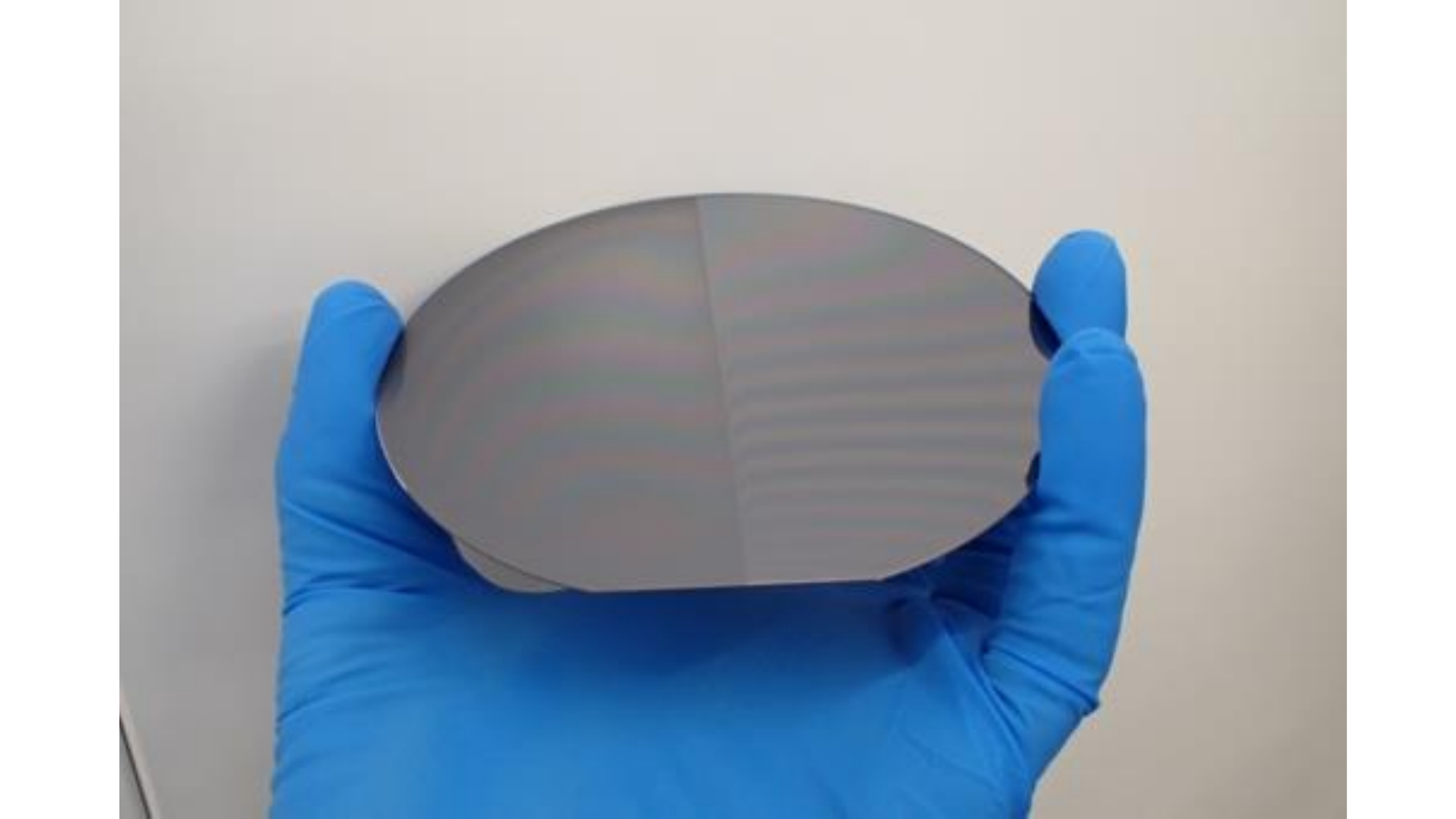}\\
\vspace{0.5em}
\includegraphics[width=3.4in]{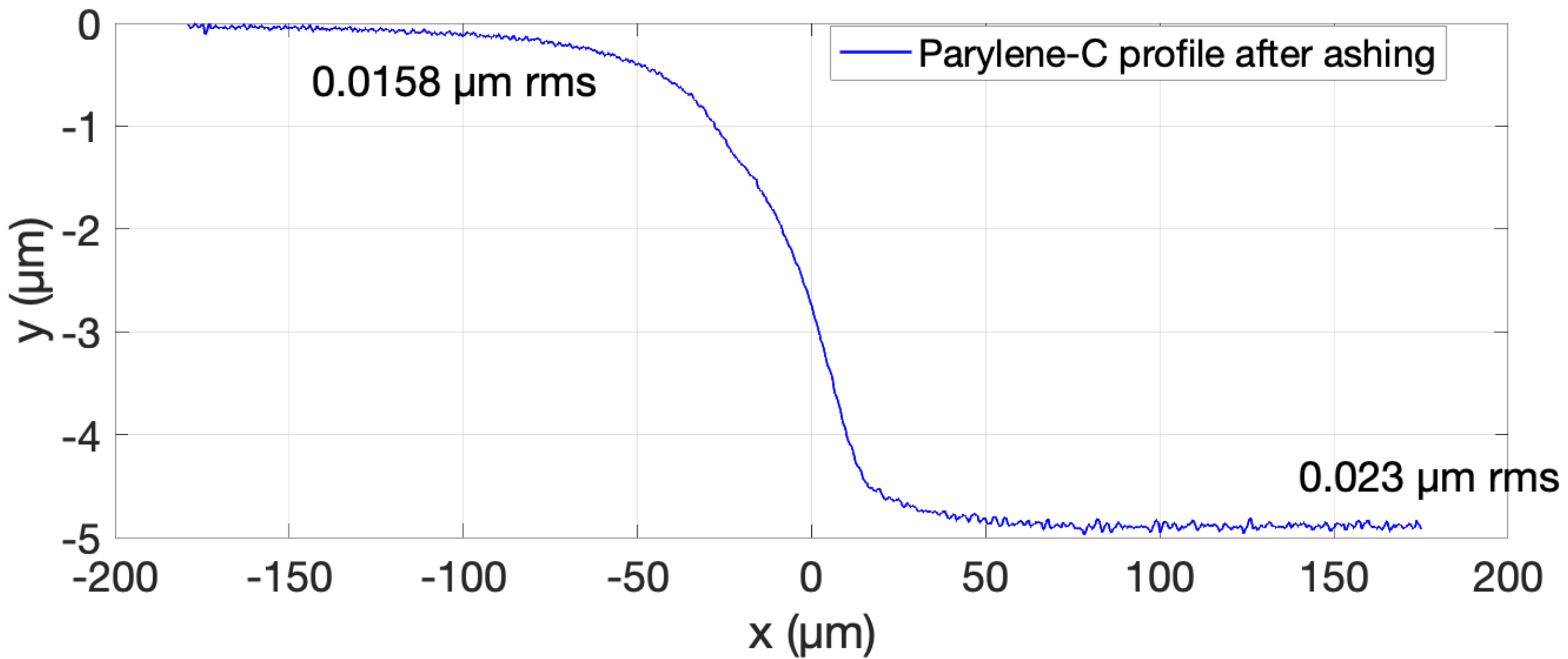}
\caption{Demonstration of two-thickness Parylene-C AR coating on a flat silicon test wafer. Top: Photograph of the test wafer with two thicknesses visible on the left and the right halves of the wafer. Bottom: Profilometry of the AR-coating thickness.}
\label{fig:parylene}
\end{figure}

In the FIRESS instrument, every detector observes at a slightly different wavelength than its neighbors due to the spectrometer's dispersion. For all four FIRESS bands, the wavelength ratio of the band edges is $\approx$ 1.8, which is larger than the bandwidth of a quarter-wave thickness AR-coating layer.  Therefore, to optimize the optical performance within each band, we implement four stepped-thickness coatings along the array length, matched to the spectrometer wavelength at that focal plane location. Once a uniform conformal coating layer is deposited, oxygen plasma etching of the coating through a shadow mask is used to make stepped gradations in the Parylene-C thickness. The oxygen plasma process etches Parylene-C isotropically because it does not depend on ion-bombardment of the surface, but rather on neutral oxygen radicals. The post-deposition plasma etching also reduces variability in the as-deposited thickness. Here, we demonstrate the process by creating a two-thickness coating on a flat silicon test wafer, as shown in Figure \ref{fig:parylene}. The profilometry shows that the transition is not a step function. In fact, the tapered transition is desirable for improved optical efficiency of the spectral pixels in this region. To demonstrate this optimization on curved lens profiles, we are currently fabricating kilo-pixel arrays with four stepped-thickness coatings.

The AR coating layer needs to be robust against cracking or delamination during thermal cycling. We tested a Band~4 kilo-pixel array with a uniform 28 \textmu m thick coating by thermally cycling it between 77 K (liquid nitrogen) and 323 K (hot plate). Even after 10 rapid thermal cycles, the AR coating showed no signs of damage, demonstrating its robustness to cryogenic cycling.

\section{Bonding}\label{sec:bonding}
The AR-coated lenslet array is aligned and bonded to the KID array using a Smart Equipment Technology FC300 flip-chip bonder\footnote{www.set-na.com/fc300/}. The flip-chip bonder provides $\pm$ 0.5 \textmu m post-bond accuracy, which is well within the $\pm$ 10 \textmu m lenslet-detector alignment tolerance for FIRESS. Using alignment marks etched on the back of the dies, a bidirectional microscope on the bonder is used to align the dies to within 3 \textmu m. The aligned dies are then bonded together with a thin layer of Epo-Tek 301 epoxy\footnote{www.epotek.com/product/301/}. To minimize the bond thickness and prevent air gaps from forming in the bond layer, continuous pressure is applied to the dies while the epoxy cures\cite{cothard2024}.

Choosing an appropriate bond force involves a trade-off between applying enough force for the epoxy to flow out and thin sufficiently before curing, versus avoiding excessive force that could cause the die to bow and trap excess epoxy along the midline of the array. We use viscous flow and elastic model simulations of the lenslet-detector array with trapped epoxy to determine the bond force required to minimize epoxy thickness without excessively bowing the dies. For FIRESS, the epoxy thicknesses (and, consequently, the bond forces) were selected such that the absorption and reflection loss from the bond layers remain below 5\%.

Similar to the optical characterization of Parylene-C described in Section \ref{sec:ar_coating}, we prepared a 77 \textmu m thick free-standing Epo-Tek 301 epoxy sample and measured its spectral transmission at 5 K using an FTS, as shown in Figure \ref{fig:fts}. While the epoxy's broad absorption feature lies below the shortest PRIMA wavelength, transmission for Band~1 is comparatively lower than that for Band~4. Using this FTS data, we calculated that the epoxy bond layer needs to be below 1~\textmu m and 6~\textmu m for Band~1 and Band~4, respectively, to meet the 5\% loss requirement for PRIMA. The thicker silicon die used for Band~1 (see Section \ref{sec:lens_design}) provides the mechanical stiffness required to withstand the higher bond forces necessary for producing thinner bond layers. 

To validate our bonding process, we performed destructive sectioning and lapping of test arrays to accurately measure the bond layer thickness. We measured bond thicknesses of 1 -- 4 \textmu m for the thinner Band~4 dies and achieved thicknesses of $<$ 1 \textmu m for the thicker Band~1 dies, meeting the PRIMA requirements. We have cryogenically tested prototype lenslet-detector arrays with the epoxy bonding layer down to $\sim$~50~mK bath temperature \cite{day2024} and plan to fully characterize the kilo-pixel arrays in flight-like packaging \cite{bradford2025}.  

\section{Summary}\label{sec:summary}
PRIMA uses monolithic kilo-pixel silicon lenslet arrays to couple incoming radiation onto KID arrays. We report on the optimizations of fabrication, lens design, AR coating, and bonding processes developed for the PRIMA FIRESS lenslet arrays. These optimizations are targeted towards increasing the instrument's optical efficiency and reducing stray light. We have refined our grayscale lithography followed by DRIE fabrication process to correct for lateral etching effect, enabling us to etch deeper and wider lenses. Using an FTS, we have characterized the far-infrared spectral transmission of Parylene-C AR coating and \mbox{Epo-Tek 301} epoxy bonding layers to optimize their thicknesses for PRIMA. Finally, we have demonstrated stepped-thickness AR coating to improve optical performance within each FIRESS band.

Once bonded, the lenslet-detector array is tested for resilience under relevant environmental stresses and for optical performance to meet PRIMA requirements. We have tested the arrays through thermal cycling and plan to vibration test them in flight-like packaging. Through destructive and non-destructive metrology, we have characterized the as-fabricated lenslet arrays. At the time of writing, the optimized lenslet arrays are being bonded to KID arrays. The optical performance of these lenslet-bonded detector arrays will be measured in a test cryostat equipped with a calibrated blackbody source.

\section*{Acknowledgments}
This work was supported by NASA GSFC Internal Research and Development (IRAD) program and NASA SAT grant (80NSSC19K0489). S.D. is supported under NASA-JHU Cooperative Agreement (80NSSC19M005).

\bibliography{references}
\bibliographystyle{IEEEtran}

\end{document}